\documentclass{wccm}

\usepackage{epsfig}
\usepackage{graphics}
\usepackage{amsmath,amsfonts}
\usepackage{overcite}
\usepackage{fancyheadings}
\usepackage{url}
\usepackage[applemac]{inputenc}

	\newcommand {\tens}[1] {\boldsymbol{#1}}

\title{A damage mechanics model for quasi brittle materials based on the structured deformation theory\vspace{6pt}}
\author{Marc Fran\c cois\\[6pt]
\fontsize{11}{13}\selectfont
LMT-Cachan\\
ENS Cachan/CNRS UMR8535/UPMC/PRES UniverSud Paris\\
61, av. du Pr{\'e}sident Wilson, F-94230 Cachan\\ France\\
e-mail: francois@lmt.ens-cachan.fr
\\[14pt]}

\begin{document}

\maketitle \noindent\textbf{Key Words:} Structured deformations, Damage, Thermodynamics of irreversible processes, Yield surface, Quasi-brittle materials \vspace{6pt}

\thispagestyle{fancy} \setlength{\headheight}{14pt}
\setlength{\topmargin}{-3 mm}
\setlength{\headsep}{3mm}
\renewcommand{\headrulewidth}{0.pt}
\lhead{ } \rhead{\footnotesize {\bf CanCNSM 2008} \\ Toronto \\
June 25-29, 2008} \cfoot{}

\begin{abstract}

The structured deformation theory is used within the thermodynamics of irreversible processes framework in order to build a damage model relevant for quasi-brittle materials.\\

The cracks are supposed smeared in the body and their shape is assumed to be sinusoidal. The convex of elasticity supposes a limitation of the thermodynamic force associated to the relative sliding of the crack lips. This limitation consists of both some equivalent to the Coulomb's friction and a cohesive force issued from the Barenblatt approach. This leads to an initial yield surface that can be expressed as the Mohr-Coulomb criterion, known to be relevant for such materials in term of initial yield surface and prediction of the crack orientation.\\

The evolution law introduces an hardening and softening function that, combined to the rest of the framework, allows the description of the stress to strain evolution. The results are identified and compared to the Kupfer's database on concrete.

\end{abstract}

\pagebreak
\setlength{\headheight}{14pt}
\setlength{\topmargin}{-9mm}
\setlength{\headsep}{12mm}
\renewcommand{\headrulewidth}{0.6pt}
\lhead{\footnotesize A damage mechanics model for quasi brittle materials} \rhead{\footnotesize
M. Fran\c cois} \cfoot{}

\section{INTRODUCTION}

Damage mechanics models are generally based on an homogenization process that takes into account elementary cells containing a micro-crack. The crack opening is introduced as a function of the crack sliding\cite{andrieux_86,halm_98} without strong micro-mechanical meaning. The structured deformation framework gives one. This opening is associated to the intrinsic crack roughness that is observed in material with granular structure, as concrete or rocks. Moreover this phenomenon is associated to the dilatancy effects, the dramatic increase of the apparent Poisson's ratio, that is of first importance for the simulation of concrete structure under severe loading (impact, seismic...).

\section{Structured deformations}

\begin{figure}[htbp]
	\begin{center}
		\epsfig{file=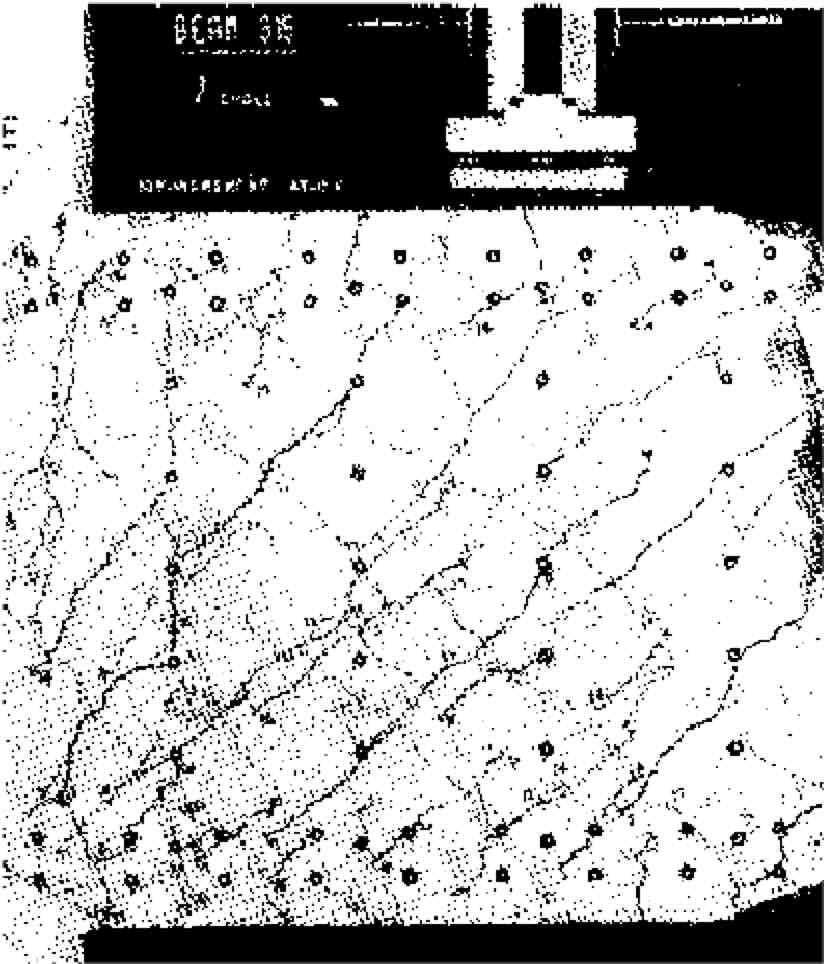,scale=0.333}
		\caption{diffuse network of microcracks in concrete}
		\label{crackedconcrete}
	\end{center}
\end{figure}

\begin{figure}[htbp]
	\begin{center}
		\epsfig{file=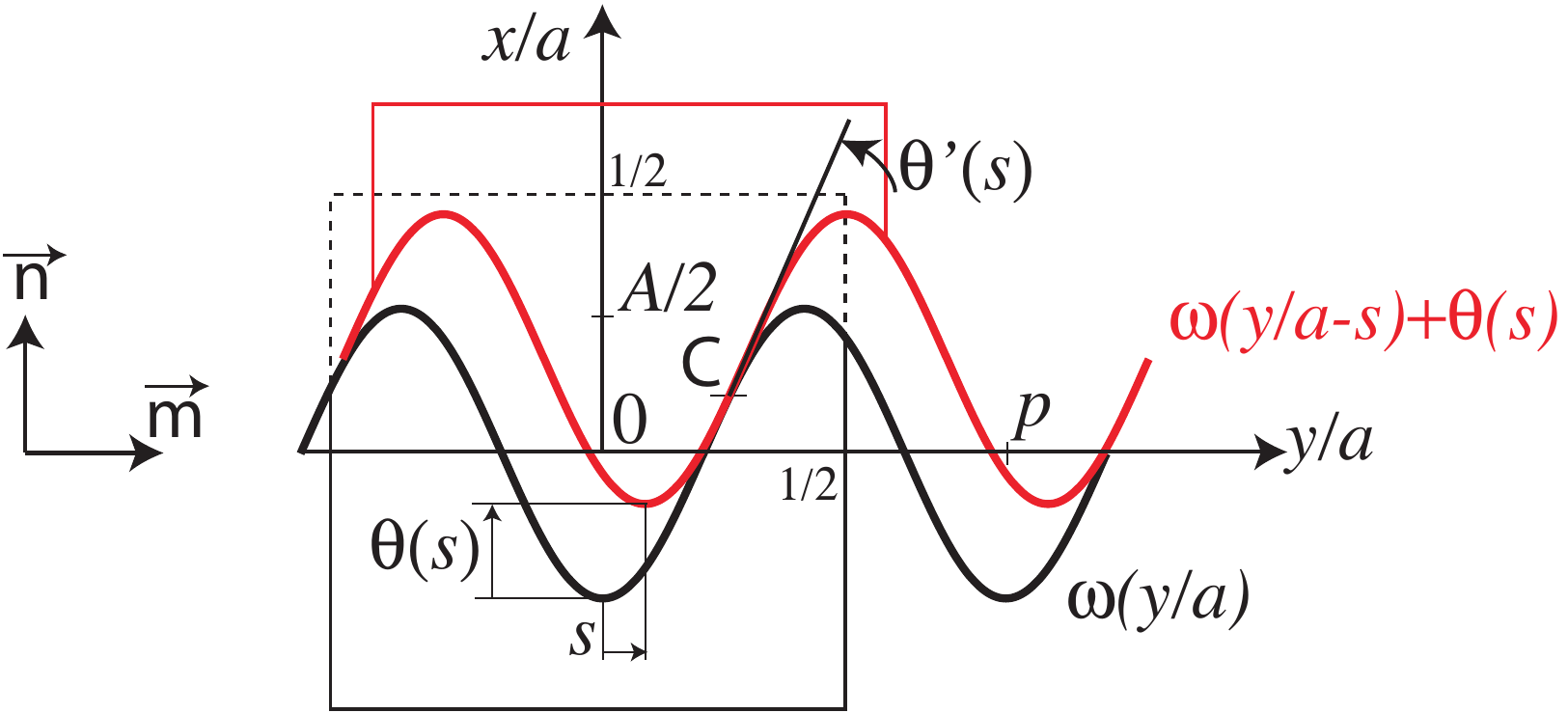,scale=0.333}
		\caption{micromechanics}
		\label{micromechanics}
	\end{center}
\end{figure}

Consistently with damage mechanics \cite{lemaitre_96} the material is supposed microcracked. Such states can be found in quasi-brittle materials. See Fig. \ref{crackedconcrete} for example. The crack orientation is supposed to vary slowly and the distance between consecutive cracks small, with respect to the stress gradient. Then we consider locally a collection of rough parallel and equidistant cracks whose normal is $\vec{n}$. The vector $\vec{m}$ (with $\vec{n}\bot\vec{m}$) is normal to the average crack plane. It represents the sliding direction of the crack.
According to Del Piero and Owen's work\cite{delpiero_93} the structured deformation is described with respect to a crack shape function $\omega(y/a)$ ($x$ and $y$ are the local coordinates along $(\vec{n},\vec{m})$). The physical length $a$ is the distance between two cracks. The crack shape is supposed to be constant along the third direction. The upper crack surface is described by $\omega(y/a-s)+\theta(s)$ (see Fig. \ref{micromechanics}) where $s$ is the crack sliding and $\theta(s)$ the induced crack opening. The contact point between the crack surfaces is C.
In the present approach the crack shape is supposed sinusoidal:
\begin{equation}
	\omega(y/a)=-\frac{A}{2}\cos{2\pi \frac{y}{ap}}.
\end{equation}
The equality of the slopes at contact point C leads to (if $s>0$):
\begin{eqnarray}
	\frac{y(C)}{a}&=&\frac{s}{2}+\frac{p}{4}+kp\quad \textrm{if }s\in]0,p/2]+kp, k\in\mathbb{Z},\\
	\frac{y(C)}{a}&=&\frac{s}{2}-\frac{p}{4}+kp\quad \textrm{if }s\in]-p/2,0]+kp, k\in\mathbb{Z}.
\end{eqnarray}
\begin{figure}[htbp]
	\begin{center}
		\epsfig{file=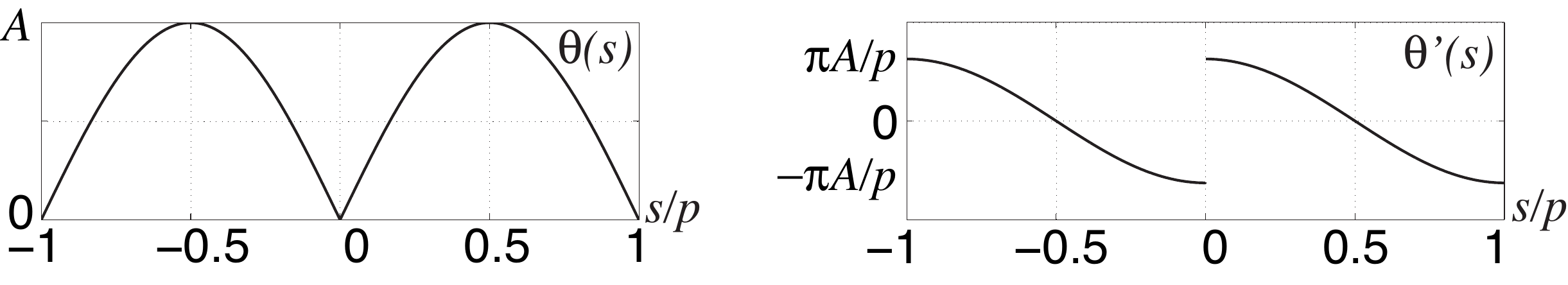,scale=0.3}
		\caption{The opening function}
		\label{theta}
	\end{center}
\end{figure}
The equality $\omega(y(C)/a)=\omega(y(C)/a-s)+\theta(s)$ gives the expression of the opening function $\theta(s)$ and it's derivative (see Fig. \ref{theta}).%
\begin{eqnarray}
	\theta(s)&=&A\left| \sin \left( \frac{\pi s}{p} \right) \right|
	\label{openingfunction}\\
	\theta'(s)&=&\frac{\pi A}{p}\,
	\mathrm{sign}\left(\sin \left( \frac{\pi s}{p} \right)\right)
	\cos \left( \frac{\pi s}{p} \right)
	\label{openingfunction}
\end{eqnarray}
This derivative presents a discontinuity for $s=kp$, $k\in\mathbb{Z}$. At these points the function reaches it's maximum absolute values and we have:
\begin{eqnarray}
	\theta'(0+kp)&\in& [-M, M]\\
	\theta'(O^++kp)=M,\, \theta'(O^-+kp) &=& -M\label{propertythetaprimzero}\\
	M &=& \frac{\pi A}{p}\\
	M &=& \max_s{\theta'(s)}
\end{eqnarray}
The homogenization process consists of the discrete derivation of the displacement vector. The continuum is seen as a collection of elementary cubes (as the one depicted on Fig. \ref{crackedconcrete}) whose side is $a$. With respect to a reference one, the displacement expresses as:
\begin{equation}
	\vec{u}(ka,la) = (ka\theta(s), kas)\quad\{k,l\}\in\mathbb{Z}
\end{equation}
The discrete gradient of the transformation is $\tens{F}$:
\begin{equation}
	\tens{F}(s) = 
	\begin{array}{|ccc|}
		\theta(s) & 0 & 0 \\
		s & 0 & 0 \\
		0 & 0 & 0
	\end{array}
\end{equation}
The strain associated to the structured deformations is the symmetric part of this gradient. 
\begin{equation}
	\tens{\varepsilon}_\mathrm{s}(s) = 
	\begin{array}{|ccc|}
		\theta(s) & s/2 & 0 \\
		s/2 & 0 & 0 \\
		0 & 0 & 0
	\end{array}
\end{equation}
At this step one can remark that the physical length $a$ is no more present in the strain (and, further, stress) equations. The remaining parameters are mathematical period $p$ and amplitude $A$. It will be convenient to use the intrinsic notation (in which $\otimes$ is the dyadic product):
\begin{equation}
	\tens{\varepsilon}_\mathrm{s}(s) = \theta(s) (\vec{n}\otimes\vec{n}) +
	\frac{s}{2}(\vec{n}\otimes\vec{m}+\vec{m}\otimes\vec{n})
	\label{structuredstrain}
\end{equation}
As in the plasticity theory, the strain $\tens{\varepsilon}$ is supposed to be the sum of it's elastic and anelastic parts (coming from the structured deformations) defined respectively by superscripts $e$ and $s$:
\begin{equation}
	\tens{\varepsilon} = \tens{\varepsilon}_\mathrm{e} + \tens{\varepsilon}_\mathrm{s}
	\label{partition}
\end{equation}
The elastic deformation $\tens{\varepsilon}_\mathrm{e}$ will be supposed independent of the crack sliding $s$. This hypothesis corresponds to neglect the stress and strain concentration around the contact points C (Fig. \ref{micromechanics}). It is possible when the crack density remains weak. An indicator of this density is $A$ (the cracks have an amplitude of $aA$ and are distant of $a$).

\section{Thermodynamics}

The present approach uses the framework of thermodynamics of irreversible processes\cite{halphen_75}. The following Helmholtz free energy is defined as the (reversible) part stored by the elasticity of the material. The (here isotropic) Hooke tensor is denoted as $\mathcal{C}$.
\begin{equation}
	2\rho\Psi(\tens{\varepsilon},s) = (\tens{\varepsilon}-\tens{\varepsilon}_\mathrm{s}):\mathcal{C}:(\tens{\varepsilon}-\tens{\varepsilon}_\mathrm{s})
\end{equation}
The internal variables (that describe the state of the material) are the elastic and anelastic strains. The associated thermodynamics force are associated to them with respect to the Helmholtz free energy. 
\begin{eqnarray}
	\tens{\sigma} &=& \frac{\partial \rho \Psi}{\partial \tens{\varepsilon}}\\
	\tens{\sigma} &=& \mathcal{C}:\tens{\varepsilon}_\mathrm{e}
\end{eqnarray}
%
The other force is $S$, that drives $s$, in other words it makes the crack slide. It's expression has a strong physical meaning as is composed of the shear stress $\sigma_\mathrm{nm}$ added with $\theta'\sigma_\mathrm{nn}$ that represents the contribution of the normal stress with respect to the slope $\theta'$ that is also the slope of the opening mechanism (see Fig. \ref{micromechanics}).
\begin{eqnarray}
	S &=& -\frac{\partial \rho \Psi}{\partial s}\\
	S &=& \theta'\tens{\sigma}:(\vec{n}\otimes\vec{n}) + 
	\frac{1}{2}\tens{\sigma}:(\vec{n}\otimes\vec{m}+\vec{m}\otimes\vec{n})\label{definitionS}\\
	S &=& \theta'\sigma_\mathrm{nn} + \sigma_\mathrm{nm}
\end{eqnarray}
The yield surface $f=0$ is defined as a bound of this thermodynamic force $S$. The resistance of the material comes from, in one hand, the Barenblatt cohesive force $B_0$ and, in the second hand, the Coulomb's friction that is supposed to be proportional to the normal stress $\sigma_\mathrm{nn}$. One can notice that the frictional aspect is considered at the macro scale as $\sigma_\mathrm{nn}$ is the normal stress to the average crack plane. A full micro approach would lead to consider the stress along the normal at the contact point C, whose angle is $\pi/2+\theta'(s)$; this may be done in further developments. 
\begin{eqnarray}
	f(\tens{\sigma},S) &=& | S | -B_0 + \varphi \sigma:\vec{n}\otimes\vec{n}\\
	f(\tens{\sigma},S) &=& | \theta'\sigma_\mathrm{nn} + \sigma_\mathrm{nm} | 
	- B_0 + \varphi \sigma_\mathrm{nn}
	\label{yieldcondition}
\end{eqnarray}
The following admissibility condition comes from the general condition $f\leq0$:
\begin{equation}
	\varphi \sigma_\mathrm{nn} \leq B_0
	\label{admissibility}
\end{equation}
The present model is an associated one, then the flux $\dot{s}$ is simply given by the normality rule that reduces here to:
\begin{eqnarray}
	\dot{s} &=& \frac{\partial f}{\partial S}\dot{\lambda}\\
	\dot{s} &=& \mathrm{sign}(S) \dot{\lambda}
	\label{normality}
\end{eqnarray}
with $\dot{\lambda}\geq0$. Then $S$ and $\dot{s}$ have the same sign and the thermal dissipation, given by the product of the forces and fluxes:
\begin{equation}
	\dot{D} = S\dot{s} \geq 0\label{dissipation}
\end{equation}
is obviously positive. When sliding occurs, the yield condition $f=0$ leads to:
\begin{eqnarray}
	\dot{D} &=& |S| \, |\dot{s}|\\
	\dot{D} &=& (B_0 - \varphi \sigma_\mathrm{nn}) \, |\dot{s}|
\end{eqnarray}
and this equation shows the equivalence of the admissibility condition \ref{admissibility} and the positiveness of the dissipation \ref{dissipation}.

\section{The induced Mohr-Coulomb criterion}

\begin{figure}[htbp]
	\begin{center}
		\epsfig{file=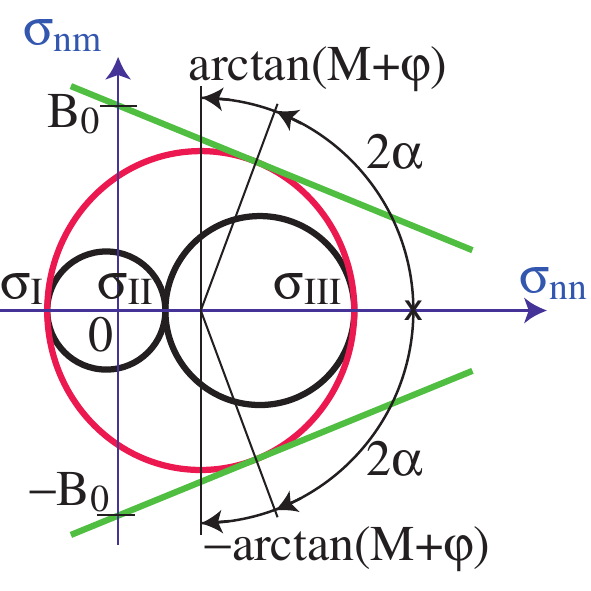,scale=0.5}
		\caption{the obtained Mohr-Coulomb criterion}
		\label{mohrcoulomb}
	\end{center}
\end{figure}
The vector $\vec{n}$ defines the crack orientation. The former crack will occur for the orientation of $\vec{n}$ that maximizes the value of the yield criterion $f$. At the initial state, $s=0$. The crack can initiate either for $\dot{s}>0$ or $\dot{s}<0$. In the first case, if $\dot{s}>0$, from Eq. \ref{propertythetaprimzero} $\theta'=M$ and, from Eq. \ref{normality}, $S>0$. In the second case, if $\dot{s}<0$, from Eq. \ref{propertythetaprimzero} $\theta'=-M$ and, from Eq. \ref{normality}, $S<0$. From Eq. \ref{yieldcondition}, the yield function writes:
\begin{eqnarray}
	\text{if }\dot{s}>0\quad f(\tens{\sigma},S) &=& 
	(M+\varphi)\sigma_\mathrm{nn} +
	\sigma_\mathrm{nm} 
	- B_0\label{condyield1}\\
	\text{if }\dot{s}<0\quad f(\tens{\sigma},S) &=&
	(M+\varphi)\sigma_\mathrm{nn} -
	\sigma_\mathrm{nm} 
	- B_0\label{condyield2}
\end{eqnarray}
At the initiation $f=0$ and this leads to the following expression of the well known Mohr-Coulomb criterion.
\begin{eqnarray}
	\sigma_\mathrm{nm}&\leq& B_0-(M+\varphi)\sigma_\mathrm{nn}\\
	\mathrm{or}\quad
	\sigma_\mathrm{nm}&\geq& -\left(B_0-(M+\varphi)\sigma_\mathrm{nn}\right)
\end{eqnarray}
The classical construction with the three Mohr circles (where $\{\sigma_{I}\leq\sigma_{I}\leq\sigma_{I}\}$ are the principal stresses), see Fig. \ref{mohrcoulomb} leads to many results: the normal $\vec{n}$ of the first crack plane is in the plane $[\vec{e}_I,\vec{e}_{III}]$ formed by the eigenvectors associated to the minimum and maximum principal stresses; the angle $\alpha=\widehat{\vec{e}_{III},\vec{n}}$ is such as:
\begin{equation}
	\alpha = \pm \left( \frac{\pi}{4}-\frac{\arctan(M+\varphi)}{2} \right)\label{alpharomeo}
\end{equation}
It can be seen that $M$ and $\varphi$ have an equivalent role for the crack initiation. The structured deformation approach lead to recover this criterion that is still one of the most accurate for quasi-brittle materials.

\section{Biaxial testing simulation}

\begin{figure}[htbp]
	\begin{center}
		\epsfig{file=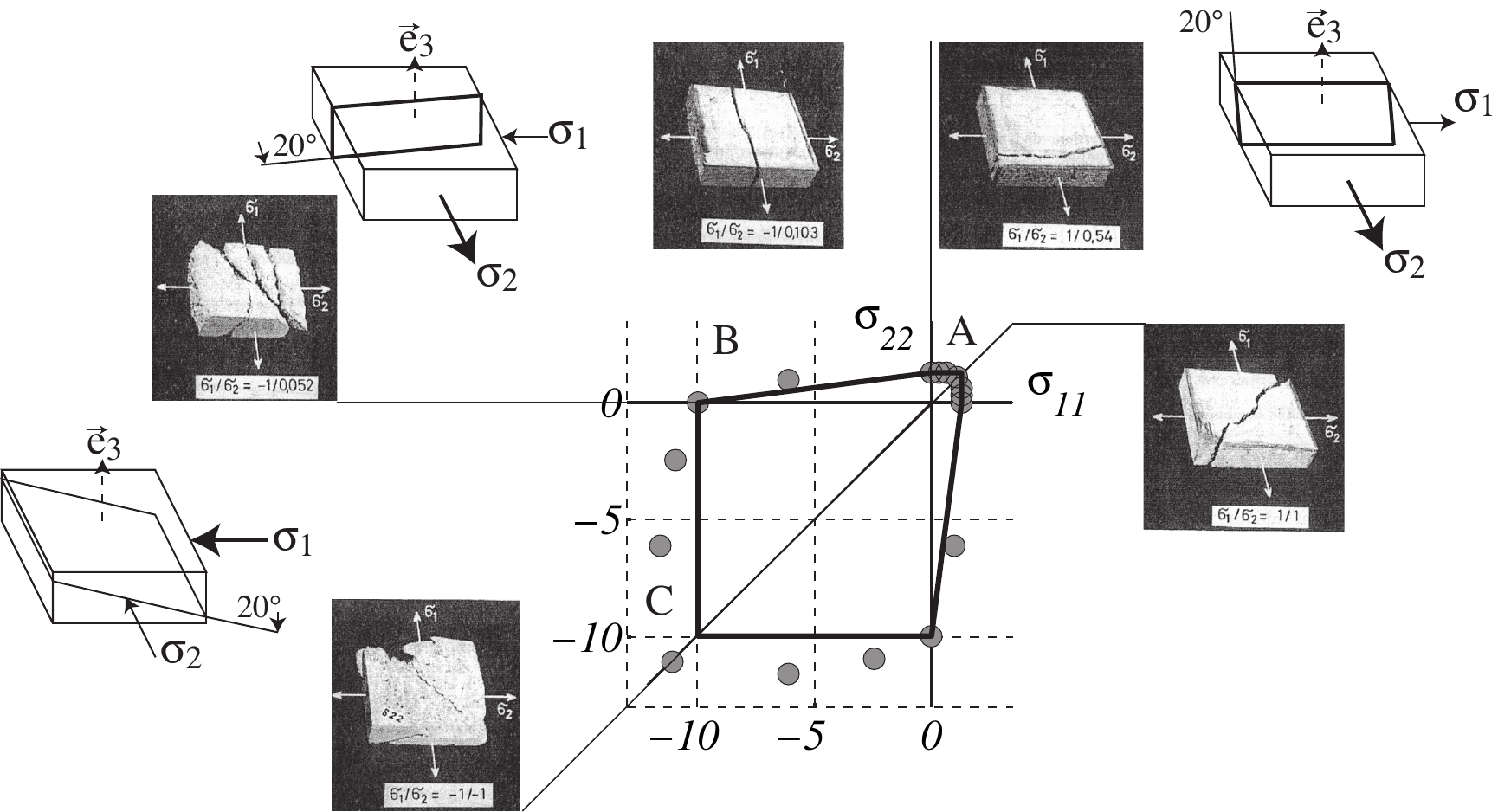,scale=0.5}
		\caption{the Mohr-Coulomb yield surface (plain), the Kupfer testings (circles) and the fracture modes of concrete}
		\label{SS_Mohr_Coulomb}
	\end{center}
\end{figure}

The Kupfer's experiments \cite{kupfer_69} are related to concrete under bi-axial states. If the best known results are the peak stress curves, he also identified the yield surface and fracture modes. With respect to the previous Mohr-Coulomb criterion, the identification on uniaxial tension and compression leads to the result presented on Fig. \ref{SS_Mohr_Coulomb} for the values $(M+\varphi)=1.21$ and $B_0=1.79$ MPa. From Eq. \ref{alpharomeo}, the angle $\alpha=19.7$ degrees. Both the yield surface and the crack modes are rather in good agreement with the experience.
%

\section{Evolution, example of tension and compression}

\begin{figure}[htbp]
	\begin{center}
		\epsfig{file=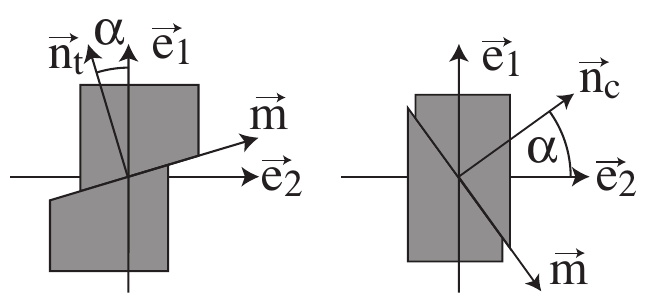,scale=0.5}
		\caption{geometrical setup: tension (left) and compression (right)}
		\label{reperes}
	\end{center}
\end{figure}

\begin{figure}[htbp]
	\begin{center}
		\epsfig{file=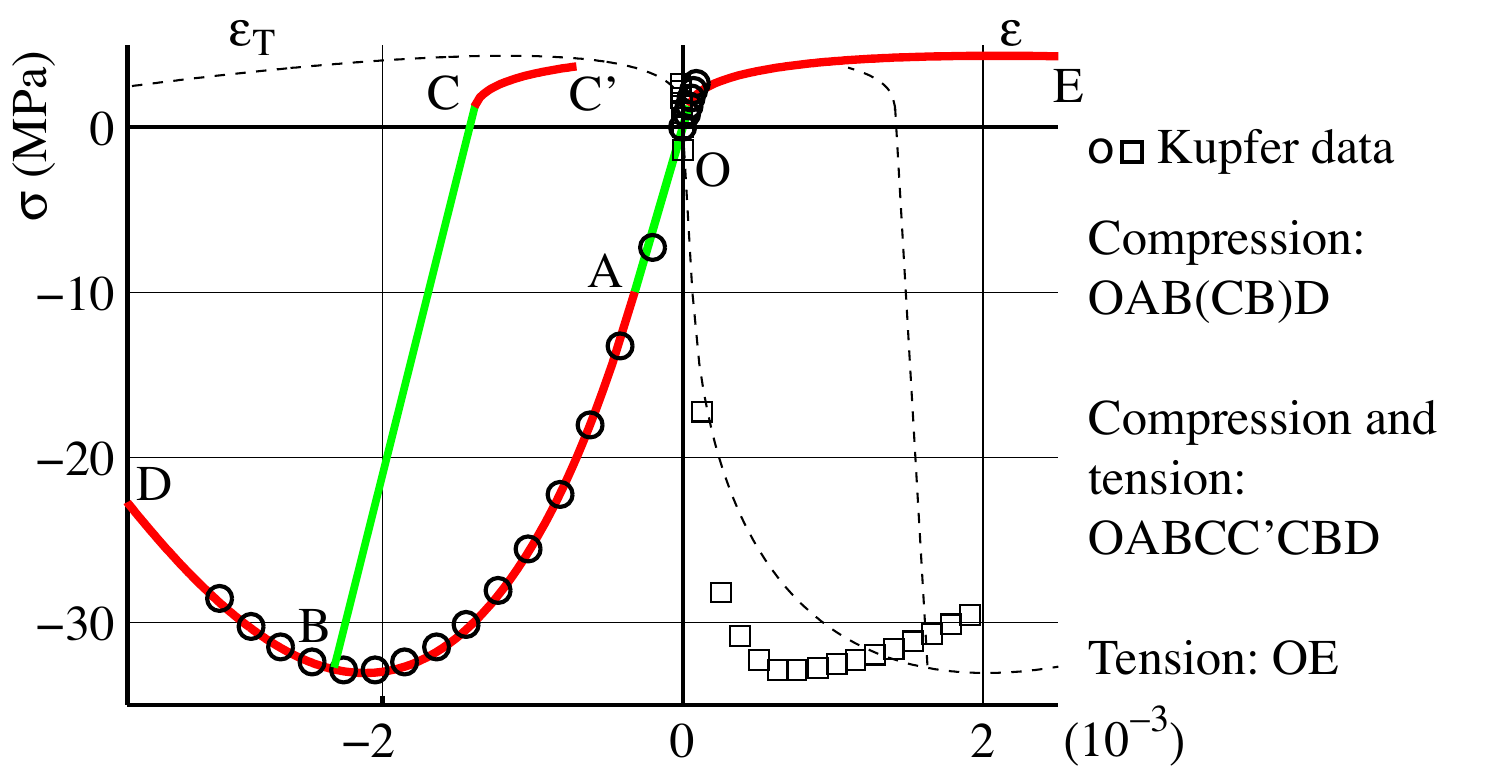,scale=0.5}
		\caption{Stress to strain graph in tension and compression.}
		\label{se}
	\end{center}
\end{figure}

The thermodynamic formalism gave all the needed equations. However it is possible to introduce a dependance of the cohesion $B$ upon the sliding $s$ (the positiveness of the dissipation only requires the positiveness of $B$), in the way of many models, for example the Dugdale-Barenblatt approach of for the cohesive zone or some zip-models. In general, one can use a function $B(s)\geq 0$ such as $B(0)=B_0$ and, if one thinks that a crack in concrete does not 'restick' after complete opening, $B(p/2)=0$. The concrete material exhibit an important hardening just after the beginning of yielding (that comes probably from the non linear hertzian contact between grains). This one is described with respect to the following hardening-softening function $B(s)$ 
\begin{eqnarray}
	B(s) &=& B_0 + B_1\sqrt{\frac{s}{p}}-B_2\frac{s}{p}\\
	B(p/2)=0 &\Rightarrow & B_2=2\left(B_0+\frac{B_1}{\sqrt{2}}\right)
\end{eqnarray}
%
%
The stress tensor is $\tens{\sigma}=\sigma \vec{e}_1\otimes\vec{e}_1$. The Mohr-Coulomb criterion involves defines the normals $\vec{n}^c$ and $\vec{n}^t$ (superscripts $^c$ and $^t$ refers respectively to compression and tension):
\begin{eqnarray}
	\vec{n}^t \in [\vec{e}_1,\vec{e}_2],&\quad& \widehat{\vec{e}_1,\vec{n}^t}=\alpha\\
	\vec{n}^c \in [\vec{e}_1,\vec{e}_2],&\quad& \widehat{\vec{e}_2,\vec{n}^c}=\alpha
\end{eqnarray}
It can be noticed that, in this case, the problem is initially transverse isotropic around $\vec{e}_1$ and that $\vec{e}_2$ and $\vec{e}_3$ have a similar role. The geometry is defined with respect to Fig. \ref{reperes} (in order to deal with $\dot{s}>0$).
In case of loading ($S>0$ and $\dot{s}>0$), the condition of yielding \ref{yieldcondition} becomes:
\begin{eqnarray}
	\sigma^t(\theta'(s)+\varphi)\cos^2(\alpha) + \sigma^t \sin(\alpha)\cos(\alpha) &=& B(s)\\
	\sigma^c(\theta'(s)+\varphi)\sin^2(\alpha) - \sigma^c \sin(\alpha)\cos(\alpha) &=& B(s),
\end{eqnarray}
and this equation gives the value of the stress during the anelastic transformation \emph{i.e.} when $s$ increases. From these expressions one can find the following simple expressions for the yield stresses in tension and compression:
\begin{equation}
	\sigma^t_y=\frac{2B_0}{\varphi+M+\sqrt{1+(\varphi+M)^2}}\quad
	\sigma^c_y=\frac{2B_0}{\varphi+M-\sqrt{1+(\varphi+M)^2}}
\end{equation}
In case of unloading ($S<0$ and $\dot{s}<0$), the condition of yielding \ref{yieldcondition} differs. But it can be checked that, when a tension follows a compression (transformation BCC' on Fig. \ref{se}), the creation of a new crack in tension (as for the virgin material) occurs before the crack closing of the already opened crack in compression. This is consistent, in first approach, with experiment.  
%
%
The strain evolution is given by the partition equation \ref{partition} and the structured strain equation \ref{structuredstrain}. We have ($E$ is the Young modulus):
\begin{eqnarray}
	\varepsilon^{t} &=& \frac{\sigma^t}{E} + \theta(s) \cos(\alpha)^2 + s \sin(\alpha)\cos(\alpha)\\
	\varepsilon^{c} &=& \frac{\sigma^c}{E} + \theta(s) \sin(\alpha)^2 - s \sin(\alpha)\cos(\alpha)
\end{eqnarray}
The identification procedure is done with respect to Kupfer's data on the remaining unknown constants ($\pi A/p + \varphi = M+\varphi=1.21$ MPa and $B_0$ have already been found).
\begin{table}[h]
\caption{material parameters (five independent constants for anelasticity)}
\begin{center}
\begin{tabular}{|c|c||c|c|c|c|c||c|c|}
$E$ (GPa) & $\nu$ & $A$ & $\varphi$ & $p$ & $B_0$ (MPa) & $B_1$ (MPa) & $B_2$ (MPa) & $M+\varphi$ \\
\hline
31.2 & 0.2 & 2.4 10$^{-3}$ & 1 & 36 10$^{-3}$ & 1.79 & 26.0 & 40.3 & 1.21\\	
\hline
\end{tabular}
\end{center}
\label{material_parameters}
\end{table}%
The compression simulation is in good agreement with experiment. In tension, the predicted post-peak seems overestimated in strain. Nevertheless, the post-peak data strongly depend upon the experimental procedure. The strain driven tests\cite{terrien_80}, such as PIED testing\cite{ramtani_90} exhibit wide post-peak behavior in tension (but not that wide). The lateral strain $\varepsilon_T=\varepsilon_{22}$ can also be deduced from Eq. \ref{partition} and Eq. \ref{structuredstrain}. We have ($\nu$ is the Poisson's ratio):
\begin{eqnarray}
	\varepsilon_T^{t} &=& -\nu \frac{\sigma}{E} + \theta \sin^2(\alpha)-s \sin(\alpha)\cos(\alpha)\\
	\varepsilon_T^{c} &=& -\nu \frac{\sigma}{E} + \theta \cos^2(\alpha)+s \sin(\alpha)\cos(\alpha)
\end{eqnarray}
The lateral strain is overestimated. On the contrary, most of classical damage model cannot exceed the value $0.5$ for the apparent Poisson's ratio (because damage theory has to respect the bound $\nu\leq0.5$ imposed by elasticity).

\section{Conclusions}

This simple damage model base on structured deformations reveals to be already relevant for quasi brittle materials, even in the present form. However many aspects remain to be checked or enhanced, for example the response under non proportional loadings for which the periodicity involved by the crack roughness in two dimensions is no more possible, due to the bi-dimensional aspect of the crack surface. This leads to further work in the structured deformation field, as the introduction of real fractal surfaces observed in such materials\cite{auradou_05}.

\bibliography{torontobiblio}

\end{document}